\documentclass[twocolumn]{aastex6}
\usepackage{textcomp}
\usepackage{amsmath}
\usepackage{graphicx}
\usepackage{caption}
\usepackage{subfigure}
\usepackage[T1]{fontenc}

\def\beq{\begin{eqnarray}}
\def\eeq{\end{eqnarray}}

\begin{document}

\title{Black hole hyperaccretion inflow-outflow model. I. long and ultra-long gamma-ray bursts}

\author{Tong Liu,\altaffilmark{1} Cui-Ying Song,\altaffilmark{1} Bing Zhang,\altaffilmark{2,3,4} Wei-Min Gu,\altaffilmark{1} and Alexander Heger\altaffilmark{5,6,7}}

\altaffiltext{1}{Department of Astronomy, Xiamen University, Xiamen, Fujian 361005, China}
\altaffiltext{2}{Department of Physics and Astronomy, University of Nevada, Las Vegas, Nevada 89154, USA}
\altaffiltext{3}{Department of Astronomy, School of Physics, Peking University, Beijing 100871, China}
\altaffiltext{4}{Kavli Institute of Astronomy and Astrophysics, Peking University, Beijing 100871, China}
\altaffiltext{5}{Monash Centre for Astrophysics, Monash University, VIC 3800, Australia}
\altaffiltext{6}{School of Physics and Astronomy, University of Minnesota, Minneapolis, MN 55455, USA}
\altaffiltext{7}{Department of Physics and Astronomy, Shanghai Jiao-Tong University, Shanghai 200240, China}
\email{tongliu@xmu.edu.cn}

\begin{abstract}
Long-duration gamma-ray bursts (LGRBs) and ultra-LGRBs (ULGRBs) originate from collapsars, in the center of which a newborn rotating stellar-mass black hole (BH) surrounded by a massive accretion disk may form. In the scenario of BH hyperaccretion inflow-outflow model and Blandford-Znajek (BZ) mechanism to trigger gamma-ray bursts (GRBs), the real accretion rate to power a BZ jet is far lower than the mass supply rate from the progenitor star. The characteristics of the progenitor stars can be constrained by GRB luminosity observations, and the results exceed usual expectations. LGRBs lasting from several seconds to tens of seconds in the rest frame may originate from solar-metallicity ($Z \sim 1~ Z_\odot$, where $Z$ and $Z_\odot$ are the metallicities of progenitor stars and the Sun), massive ($M \gtrsim 34 ~M_\odot$, where $M$ and $M_\odot$ are the masses of progenitor stars and the Sun) stars or some zero-metallicity ($Z \sim 0$) stars. A fraction of low-metallicity ($Z \lesssim 10^{-2}~Z_\odot$) stars, including Population III stars, can produce ULGRBs such as GRB 111209A. The fraction of LGRBs lasting less than tens of seconds in the rest frame is more than 40$\%$, which cannot conform to the fraction of the demanded type of progenitor star. It possibly implies that the activity timescale of central engine may be much longer than the observed timescale of prompt emission phase, as indicated by X-ray late-time activities. Alternatively, LGRBs and ULGRBs may be powered by a millisecond magnetar central engine.
\end{abstract}


\section{Introduction}

Mounting evidence suggests that long-duration gamma-ray bursts (LGRBs) originate from collapses of massive stars in star-forming and low-metallicity regions of star-forming galaxies \citep[see reviews by][]{Woosley2006,Kumar2015}. Observationally they are unambiguously associated with core-collapse supernovae (SNe) and linked to the deaths of massive stars. Now the collapsar model \citep[see e.g.,][]{Woosley1993,MacFadyen1999,Woosley2002,Zhang2004,Woosley2012} is generally acknowledged to explain the origin of LGRBs.

A black hole (BH) or a neutron star (NS) will be born in the center of a massive star right after it begins to collapse. For the BH case, the fall back matter triggers the BH hyperaccretion processes to power a relativistic jet breaking out from the envelope via neutrino-antineutrino annihilation mechanism liberating the gravitational energy of the BH \citep[e.g.,][]{Ruffert1997,Rosswog2003}, which corresponds to neutrino-dominated accretion flows \citep{Popham1999,Di Matteo2002,Gu2006,Liu2007,Kawanaka2007,Zalamea2011,Xue2013,Liu2016,Song2016}, or the Blandford-Znajek (BZ) mechanism tapping the rotational energy of the BH \citep[e.g.,][]{Blandford1977,Lee2000a,Lee2000b,Wu2013,Lei2013,Lei2017,Liu2015}. For a recent review on GRB NDAFs, see \cite{Liu2017}.
For the NS case, the spin down of a NS with a millisecond rotation period and a strong magnetic field (millisecond magnetar) extracting the rotational energy by electromagnetic torques can produce LGRBs, even super-luminous SNe \citep[e.g.,][]{Duncan1992,Usov1992,Dai1998a,Dai1998b,Kluzniak1998,Zhang2001,Dai2006,Metzger2011,Metzger2015,Lv2014,Lv2015}. Actually, hyperaccreting BHs and millisecond magnetars are the main plausible candidates for the central engine of gamma-ray bursts (GRBs).

The progenitor and central engine of ultra-LGRBs (ULGRBs) remain a mystery. The major challenge for theoretical models comes from the durations of ULGRBs. A most well-acknowledged example of ULGRBs is GRB 111209A \citep[e.g.,][]{Gendre2013,Levan2014,Zhang2014}. The discovery of super-luminous SN 2011kl following GRB 111209A \citep{Greiner2015} further enhanced the difficulty of the interpretation. Some believe that they are different from other LGRBs. Proposed models range from a blue supergiant or Population III (Pop III) progenitor star to a magnetar central engine to tidal disruption \citep{Gendre2013,Levan2014,Zhang2014,Greiner2015,Ioka2016}.

In this paper, we consider that both LGRBs and ULGRBs likely originate from BH hyperaccretion processes in the massive collapsars to investigate what types progenitor stars can power them. This paper is organised as follows. In Section 2, we propose our central engine model. In Section 3, the total jet energies and timescales of different LGRBs and ULGRBs are presented. The main results on progenitor star constraints are shown in Section 4. Conclusions and discussion are in Section 5.

\section{Model}

\begin{figure}
\includegraphics[angle=0,scale=0.4]{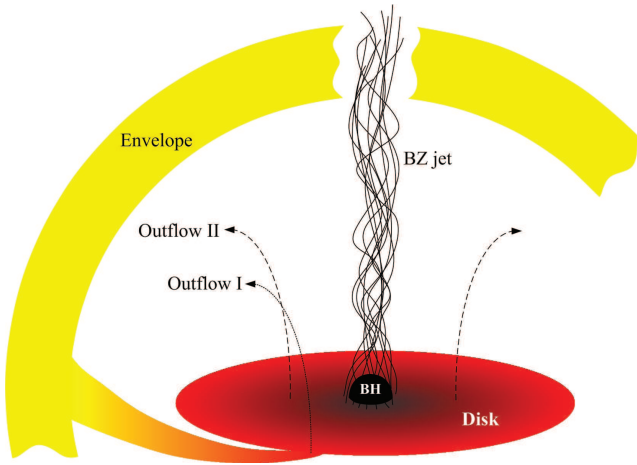}
\caption{Schematic picture of BH hyperaccretion inflow-outflow model for LGRBs and ULGRBs.}
\label{fig1}
\end{figure}

One widely discussed picture of the central engine of LGRBs and ULGRBs is shown in Figure 1. After a massive progenitor star collapses, a stellar mass BH is born in the center \citep[e.g.,][]{Heger2003}. The materials from the envelope fall back toward the BH and an accretion disk forms. A jet produced by the BZ mechanism \citep{Blandford1977} or the neutrino-antineutrino annihilation process is launched and breaks out from the envelope. If it lasts long enough, an observable LGRB or ULGRB is triggered.

Due to angular momentum redistribution, an outflow, termed as Outflow~I, is launched when the matter of envelope falls onto the outer boundary of the disk. Additionally, a strong outflow from the disk, which we shall refer to Outflow~II, has been found in theoretical models \citep[e.g.,][]{Liu2008,Gu2015}, numerical simulations \citep[e.g.,][]{Yuan2014,Jiang2014,Sadowski2015}, and observations \citep[e.g.,][]{Wang2013,Cheung2016,Parker2017}. As a result, only a few percent of the supplied mass is eventually accreted into the BH. Here we define the dimension-less factor $\lambda$, the ratio of the accretion rate at the outer boundary ($\dot M_{\rm outer}$) of the disk to the mass supply rate from the envelope ($\dot M_{\rm pro}$), to parameterize the effect of Outflow I, i.e.,
\beq
\dot{M}_{\rm outer}=\lambda\dot{M}_{\rm pro},
\label{eq:Outflow-I}
\eeq
and use a power law model to relate the accretion rate at the inner radius of the BH disk ($\dot M_{\rm inner}$) and $\dot M_{\rm outer}$ as an effort of delineating the effect of Outflow II \citep[e.g.,][]{Blandford1999,Yuan2012,Yuan2014,Sadowski2015}, which can be described by
\beq
\dot{M}_{\rm inner}=\dot{M}_{\rm outer} \left(\frac{r_{\rm inner}}{r_{\rm outer}} \right)^{p}.
\label{eq:Outflow-II}
\eeq
where $r_{\rm inner}$ and $r_{\rm outer}$ are the inner and outer boundaries of the disk, respectively, and $p$ is the index parameter. We take $r_{\rm inner}\simeq r_{\rm ms}=(3+Z_{2}-\sqrt{(3-Z_{1})(3+Z_{1}+2Z_{2})})r_{\rm g}$, and $r_{\rm outer}=100 r_{\rm g}$. Here $r_{\rm g}=GM_{\rm BH}/c^{2}$ is Schwarzschild radius, $M_{\rm BH}$ is the mass of the BH, $r_{\rm ms}$ is the dimensionless marginally stable orbit radius of the disk \citep[e.g.,][]{Bardeen1972,Kato2008}, $Z_{1}=1+(1-a_*^{2})^{1/3}[(1+a_*)^{1/3}+(1-a_*)^{1/3}]$, and $Z_{2}=\sqrt{3a_*^{2}+Z_{1}^{2}}$ for $0 < a_* <1$, where $a_*$ is the dimensionless spin parameter of the BH.

One can see that the effect of Outflow I is parameterized with the parameter $0 < \lambda < 1$, and the effect of Outflow II is delineated through the index parameter $p$. Outflow I appears near the outer boundary of the disk, which results from the difference between the angular momentum of the progenitor star and that of the outer boundary of the disk. The parameter $\lambda$ reflects how much matter from the progenitor stars turns into the matter of the disks. Once the disk forms, Outflow II will naturally emerge. The parameter $p$ represents the strength and the radial distribution of the disk outflows. In our calculation, $p=0.8$ is adopted \citep[e.g.,][]{Yuan2012,Yuan2014,Sadowski2015}, which indicates that very strong disk outflows are produced.

Moreover, for the same BH spin parameter and accretion rate, the BZ luminosity is larger by about two orders of magnitude than neutrino annihilation luminosity \citep[e.g.,][]{Kawanaka2013,Liu2015,Lei2017}. Once considering that two mechanisms have the same conversion efficiency to power a certain GRB, the values of the BH spin parameter or the accretion rate for the BZ mechanism can be lower than those for the neutrino annihilation mechanism. This suggests that the BZ mechanism is favored to power GRBs with long activity durations. Moreover, considering the strong outflow from the disk, the inner accretion rate is essentially always lower than the ignition accretion rate of NDAFs \citep[for $a_* =0.95$ and the viscosity parameter $\alpha =0.1$, the ignition accretion rate is about 0.021 $M_\odot~\rm s^{-1}$, where $M_\odot$ is the mass of Sun, see e.g.,][]{Chen2007,Zalamea2011,Liu2017}.

Since it has been shown that the BZ mechanism is more effective than the neutrino annihilation processes to power a relativistic jets, we assume that the jet is driven by the BZ mechanism, which is connected with $\dot M_{\rm inner}$ \citep{Blandford1977,Lee2000a,Lee2000b}:
\beq
\dot{E}_{\rm BZ}=1.7\times10^{20}a_*^{2}m^{2}B_{\rm inner, G}^{2}F(a_*)~{\rm erg~s^{-1}},
\eeq
where $B_{\rm inner, G}=B_{\rm inner}/1~\rm G$ is the dimensionless magnetic strength at the inner boundary of the disk, $m=M_{\rm BH}/M_{\odot}$, and
\beq
F(a_*)=[(1+q^{2})/q^{2}][(q+1/q)\arctan(q)-1]
\eeq
is a spin-dependent dimensionless parameter, and $q=a_*/(1+\sqrt{1-a_*^{2}})$. According to the balance between the ram pressure of the innermost part of the disk $P_{\rm inner}$ and the magnetic pressure on the BH horizon, we derive
\beq
\frac{B_{\rm inner}^{2}}{8\pi}=P_{\rm inner} \sim \rho_{\rm inner} c^{2}\sim\frac{\dot{M}_{\rm inner}c}{4\pi r_{\rm H}^{2}},
\eeq
where $r_{\rm H}=(1+\sqrt{1-a_*^{2}})r_{\rm g}$ denotes the radius of the BH horizon, $\dot{M}_{\rm inner}$ and $\rho_{\rm inner}$ denote the net accretion rate and density at the inner boundary of the disk.

The magnetic field strength threading the BH horizon can be then estimated by
\beq
B_{\rm inner}\simeq 7.4\times10^{16}  \dot{m}_{\rm inner}^{1/2} m^{-1}(1+\sqrt{1-a_*^{2}})^{-1}~ \rm G.
\eeq
where $\dot{m}_{\rm inner}=\dot{M}_{\rm inner}/(M_{\odot}~\rm s^{-1})$. Inserting this equation into Equation (3), the BZ jet power can be rewritten as
\beq
\dot{E}_{\rm BZ}=9.3\times10^{53}a_*^{2}\dot{m}_{\rm inner} F(a_*)(1+\sqrt{1-a_*^{2}})^{-2}~{\rm erg~s^{-1}}
\label{eq:BZ}
\eeq

Since it takes a relatively long time scale (at least $\sim 10$ s) for the jet to break out from the progenitor star, we take $a_*\approx0.86$ in our calculations because this value of $a_*$ is the asymptotic value of the spin evolution of a BH surrounded by a Keplerian accretion disk \citep[e.g.,][]{Song2015,Lei2017}.

On the other hand, the BZ power equals to the total mean jet luminosity $L_{\rm j}$ of the GRB, which includes the radiated $\gamma$-ray power in the prompt emission phase and the kinetic energy power of the outflow in the afterglow phase, i.e.,
\beq
\dot{E}_{\rm BZ}=L_{\rm j}.
\eeq

\begin{figure}
\includegraphics[angle=-90,scale=0.3]{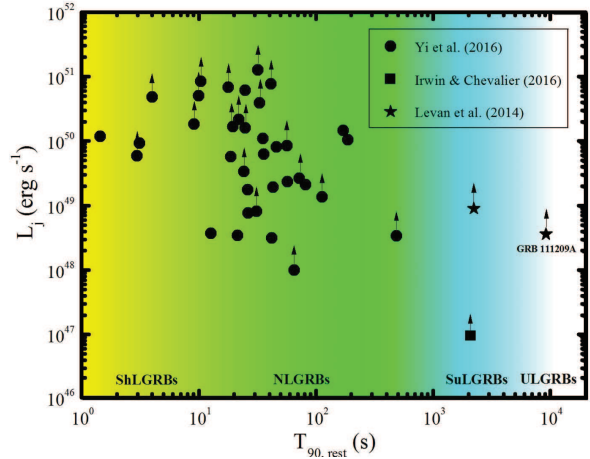}
\caption{Beaming corrected jet luminosity $L_{\rm j}$ versus duration $T_{\rm 90,rest}$ of a sample of LGRBs. The black filled circles, squares and stars denote data collected from different references. Arrows denote lower limits of jet luminosity. Different colors denote different subclasses of LGRBs defined in this paper.}
\label{fig2}
\end{figure}

Additionally, we can estimate $\dot{M}_{\rm pro}$ with the data of pre-SN models \citep[e.g.,][]{Suwa2011,Woosley2012,Matsumoto2015}, i.e.,
\beq
\dot{M}_{\rm pro}= \frac{d M_{\rm r}}{dt_{\rm ff}}=\frac{dM_{\rm r}/dr}{dt_{\rm ff}/dr}=\frac{2M_{\rm r}}{t_{\rm ff}(r)} \left(\frac{\rho}{\bar{\rho}-\rho} \right),
\eeq
where $\rho$ is the mass density of the progenitor star, $\bar{\rho}=3M_{r}/(4\pi r^{3})$ is the mean density within radius $r$, and $M_{\rm r}$ is the mass coordinate. The free-fall timescale $t_{\rm ff}$ can be calculated from
\beq
t_{\rm ff}(r)=\sqrt{\frac{3\pi}{32G\bar{\rho}}}=\frac{\pi}{2}\sqrt{\frac{r^{3}}{2G M_{\rm r}}}.
\eeq

Considering the pressure balance at the progenitor envelope and the interface of the jet, we acquire the velocity of the jet head \citep[e.g.,][]{Matzner2003}
\beq
\beta_{\rm h}=\frac{1}{1+\widetilde{L}^{-1/2}},
\eeq
where
\beq
\widetilde{L}\equiv\frac{L_{\rm j}(t-r_{\rm h}/c)}{\pi\theta_{\rm j}^{2}r_{\rm h}^{2}\rho(r_{\rm h})c^3},
\eeq
and $\theta_{\rm j}$ is the jet half-opening angle. The radius of jet head can be obtained from
\beq
r_{\rm h}(t)=\int^{t}_{0}c\beta_{\rm h}dt.
\eeq
The jet breakout time $t_{\rm bo}$ is defined by $r_{\rm h}(t_{\rm bo})=r_{*}$, where $r_{*}$ is the boundary of the progenitor star. We assume that the jet is launched when the BH mass reaches 3$M_\odot$ and set $t=0$ at this time. In other words, the enclosed mass within radius $r_{0}$ is 3$M_\odot$, i.e., $M_{r_{0}}=3 M_{\odot}$, $t=t_{\rm ff}(r)-t_{\rm ff}(r_{0})$. Of course the BH mass just keeps growing in the accretion processes. We assume that the value of the BH mass is a constant after the jet breaks out because the mass supply rate becomes much lower than that in the start of the accretion phase.

Once a jet breaks out from the star, an observable LGRB finally emerges, which can trigger the observation instruments to record this event. It starts from $t_{\rm bo}$ and lasts for about $T_{{90}, \rm rest}$ \citep[e.g.,][the rest-frame duration can be expressed by $T_{{90}, \rm rest} = T_{90}/(1+z)$, where $z$ is the redshift]{Bromberg2012}. Using the beaming corrected mean luminosity (defined as beaming-corrected prompt $\gamma$-ray energy and afterglow kinetic energy divided by $T_{{90}, \rm rest}$ \citep[e.g.,][]{Yi2017}, one can derive the mean accretion rate at the inner boundary of the disk through the BZ power formula (Equation \ref{eq:BZ}). Then the characteristics of the progenitor star can be inversely constrained by using Equations \ref{eq:Outflow-I} and \ref{eq:Outflow-II}.

\section{GRB data}

\begin{figure}
\includegraphics[angle=0,scale=0.75]{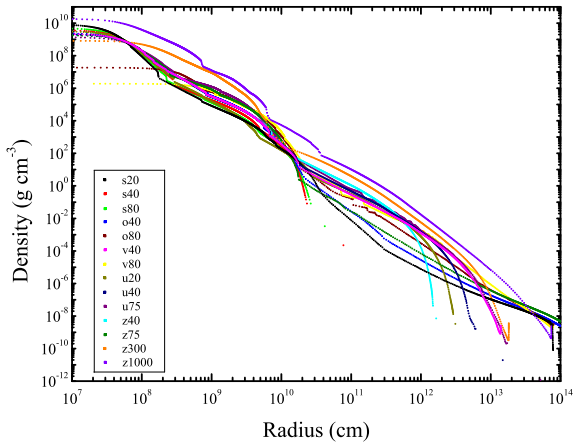}
\caption{Density profiles of the progenitor stars with different masses and metallicities.}
\label{fig3}
\end{figure}

In our analysis, the observational data are collected to derive the beaming-corrected jet luminosity $L_{\rm j}$ and time scale $T_{\rm 90,rest}$ of LGRBs and ULGRBs.

In Figure 2, the majority of data are taken from \citet{Yi2017}, who have carefully analyzed the beaming corrected jet luminosity by properly treating the prompt emission data, afterglow data, and the jet break data. These dots are denoted in black filled circles. Squares and stars denote the data collected by \citet{Irwin2016} and \citet{Levan2014}, respectively. The arrows denote the GRBs with the lower limit of jet luminosity (due to the lack of detection of a jet break).

\begin{figure}
\includegraphics[angle=0,scale=0.91]{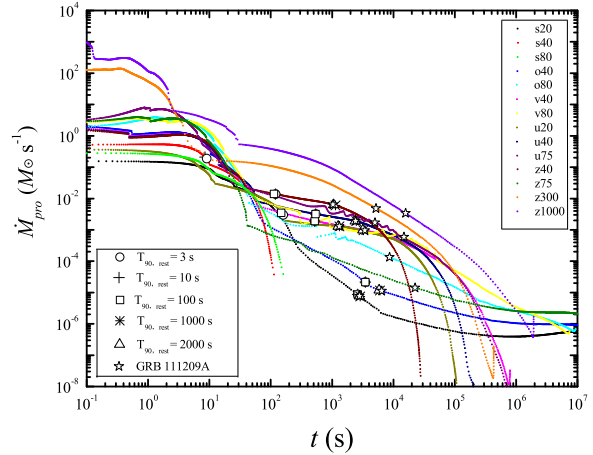}
\caption{Mass supply rates of the progenitor stars with different masses and metallicities.}
\label{fig4}
\end{figure}

\begin{figure*}
\centering
\includegraphics[angle=0,scale=1.8]{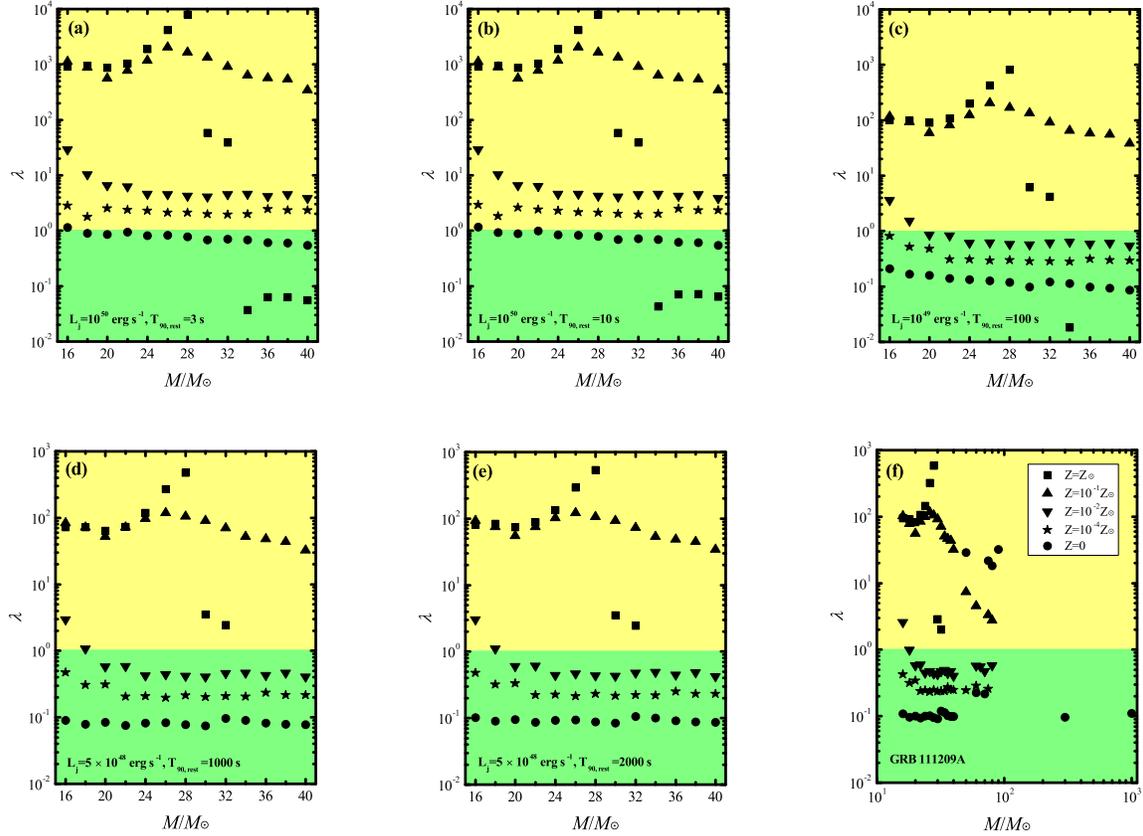}
\caption{Constraints on the progenitor stars of LGRBs and ULGRBs with different masses and metallicities.}
\label{fig5}
\end{figure*}

As shown in the figure, the typical jet luminosities $L_{\rm j}$ of LGRBs lasting less than ten seconds, about several hundred seconds, and about several thousand seconds are about $10^{50}~\rm erg~s^{-1}$, $10^{49}~\rm erg~s^{-1}$, $5 \times 10^{48}~\rm erg~s^{-1}$, respectively. For convenience of the following discussion, one can roughly divide LGRBs into four regimes, which are denoted by different colors in the figure. We define these regimes of LGRBs based on the rest frame durations: short-LGRBs (ShLGRBs, less than about ten seconds), normal-LGRBs (NLGRBs, about ten to one thousand seconds), and super-LGRBs (SuLGRBs, between one thousand and ten thousand seconds). In addition, GRBs lasting longer than ten thousand seconds are usually defined as ULGRBs. Notice that the definition of these regimes is phenomenological and arbitrary, which is convenient for us to discuss their typical $L_{\rm j}$ to constrain the characteristics of the progenitor stars as shown in Figure 5.

As a typical ULGRB, GRB 111209A at $z=0.677$ was detected by $Swift$ \citep{Hoversten2011} and continuously observed by Konus-WIND \citep{Golenetskii2011}. Its total isotropic energy output \citep{Nakauchi2013} and duration are $E_{\rm iso} \approx 1.54 \times 10^{54}$ erg and $T_{90} \approx 15000$ s. A super-luminous SN 2011kl was detected to be associated with it \citep{Greiner2015}. It is possible to place a lower limit on the jet opening angle \citep{Levan2014} of $\theta_{\rm j}> 0.21$. These values are used to estimate the jet luminosity of the burst. Moreover, we set $\theta_{\rm j}$ as 0.1 for ShLGRBs and NLGRBs, and 0.2 for SuLGRBs in Equation (12), if a jet angle is not directly measured from the data.

\section{Results of progenitor star constraints}

In order to constrain the characteristics of the progenitor stars of LGRBs and ULGRBs, the density profiles of stars with different masses and metallicities should be given first, which are provided by coauthor A.H. and displayed in Figure 3. The signs s, o, v, u, z represent the metallicity values $Z =10^{-1}~Z_\odot$, $10^{-2}~Z_\odot$, $10^{-4}~Z_\odot$, and 0, respectively \citep{Woosley2002}, where $Z$ and $Z_\odot$ are the metallicities of progenitor stars and the Sun. According to Equations (9) and (10), we can calculate the mass supply rates $\dot{M}_{\rm pro}$ of different progenitor stars with different masses and metallicities, as shown in Figure 4. The significant differences of $\dot{M}_{\rm pro}$ for different stars mainly come from the different density of the stars. The symbol on each curve represents the jet breakout time of that particular star estimated by Equations (11)-(13).

In the above scenario, by using the observational data of LGRBs and ULGRBs to define the BZ power, one can then place a constraint on the properties of GRB progenitor stars, including their masses and metallicities. If the required parameter $\lambda$ exceeds unity for a certain type of star, this star is ruled out as the progenitor star of that particular GRB. GRBs with different durations require different amount of masses to be accreted into the BH, and therefore pose different constraints on the properties of progenitor stars.

Figure 5 presents the required $\lambda$ values for progenitor stars with different masses and different metallicities. The $\lambda=1$ line separates the allowed (green) and disallowed (yellow) regions. For each sub-type of LGRBs, the masses of the potential progenitor stars are denoted in symbols with a mass interval $\sim 2~M_\odot$. The mass range spans from 16 to 40 $M_\odot$ in Figure 5 (a)-(e), but some symbols are clearly missing, which means that no corresponding LGRB can be produced by the star with the relevant mass and metallicity for any value of $\lambda$. One can draw the following conclusions from Figure 5: ShLGRBs can be produced by solar-metallicity ($Z \sim 1~ Z_\odot$), massive ($M \gtrsim 34 ~M_\odot$, where $M$ is the masses of the progenitor star) stars or some zero-metallicity ($Z \sim 0$) stars.

For NLGRBs and SuLGRBs, most low metallicity ($Z \lesssim 10^{-2}~ Z_\odot$) stars are favorable, and the solar-metallicity massive stars with $M \sim 34 ~M_\odot$ are not ruled out only for LGRBs lasting 100 s.

For ULGRBs, we use the isotropic energy and timescale of GRB 111209A to constrain the progenitor as shown in Figure 5 (f). One can see that only low-metallicity ($Z \lesssim 10^{-2}~Z_\odot$) stars with $M \gtrsim 20 ~M_\odot$, including population III (Pop III) stars, can produce ULGRBs. Contrary to intuition, some zero-metallicity stars with tens of solar mass cannot trigger ULGRBs, since their density profiles cannot bear accretion lasting for more than ten thousand seconds after the jet breaks out their envelopes. In our calculations, most of progenitor stars of NLGRBs and SuLGRBs can also produce ULGRBs. The reasons include: (a) the high efficiency of the BZ mechanism requires a low accretion rate; (b) the low density ($\lesssim 10^{-7} ~\rm g~cm^{-3}$) at the outer envelope of the progenitor star is considered, so the accretion timescale is satisfied for ULGRBs. This result is quite different from the previous conclusions that ULGRBs demand a metal poor blue supergiant with mass greater than $\sim 70~M_\odot$ \citep[e.g.,][]{Nakauchi2013,Kashiyama2013}. In these works, the detailed descriptions on the progenitors of ULGRBs are modelled. They also discussed the jet-cocoon formation and evolution before and after the jet breaking out. Since the low efficiency of the jet power, $\sim 10^{-4}$, is adopted, and the effective stellar surface is defined at the radius with the density, $\sim 10^{-7} ~\rm g~cm^{-3}$, the requirement on the star mass is more stringent than that in our results.

The rest-frame duration of LGRBs as defined as $T_{{90}, \rm rest}$ has a distribution peaking at about ten seconds \citep[e.g.,][]{Zhang2013,Zhang2014}. The number ratio of ShLGRBs to LGRBs is more than 40$\%$. However, based on the demanded progenitor properties, ShLGRBs should be much rarer than NLGRBs. We consider that most of ShLGRBs have an intrinsic duration much longer than a few seconds, due to the so-called ``tip-of-iceberg'' effect \citep{Lu2014,Li2016}. This is consistent with the observations of early X-ray afterglows that show extended central engine activities that define an effective burst duration peaking at a few hundred seconds \citep[e.g.,][]{Burrows2005,Liang2006,Zhang2006a,Luo2013,Zhang2014,Mu2016}.

Furthermore, we have tested how sensitive our results depend on BH mass, the disk outer boundary radius and outflow index $p$. For the reasonable values of these parameters obtained from numerical simulations, our results are insensitive to these parameters.

\section{Conclusions and discussion}

In the framework of the BH hyperaccretion inflow-outflow with a BZ jet, the characteristics of the progenitor stars of LGRBs and ULGRBs are tightly constrained. First, ShLGRBs may originate from solar-metallicity ($Z \sim Z_\odot$), massive ($M \gtrsim 34 ~M_\odot$) stars or some zero-metallicity ($Z \sim 0$) stars. It provides an apparent contradiction between the observational facts and the model predicted progenitor types, suggesting that the true duration of the burst is actually longer than $T_{90}$. This is consistent with the X-ray afterglow observations \citep[e.g.,][]{Burrows2005,Liang2006,Zhang2006a,Luo2013,Zhang2014,Mu2016}. Alternatively, the magnetar central engine may be at play. For ULGRBs, our model suggests that only a only a small fraction of low-metallicity ($Z \lesssim 10^{-2}~Z_\odot$) stars, including Pop III stars, are able to produce ULGRBs like GRB 111209A. This is quite different from the previous theoretical results on ULGRBs.

In our model, the angular momentum distributions \citep[e.g.,][]{Fryer2005,WoosleyH2006} of the progenitor stars is not considered. This may partially affect out results. Nonetheless, our results are approximately valid for slowly rotating stars. Another effect of rotation is that it would change the mechanical and thermal equilibrium of the star, making the star hotter at the poles and cooler at the equator. Since the mass supply for the accretion disk is provided from the equatorial direction of the star, the anisotropic temperature distribution inside the star would result in a series of consequences on mass loss rate (e.g., may stripe the hydrogen envelope), circulation current, evolution and lifetime of the chemical abundance, magnetic flux, and mass density \citep[e.g.,][]{Heger2005,Yoon2005,Barkov2010,Maeder2012}. Furthermore, internal differential rotation may generate instabilities and mixing \citep[e.g.,][]{Meynet2005,Maeder2012}. These effects should be studied in detail in the future to place better constraints on the progenitor stars. In any case, Outflow I would be significantly enhanced with the rotation effect included, resulting in an even smaller $\lambda$, especially for rapidly rotating stars. As a result, the constraints on the progenitor stars become more demanding.

Furthermore, it is noteworthy that the central engine of at least some LGRBs might be a millisecond magnetar \citep[e.g.,][]{Duncan1992,Usov1992,Dai1998a,Dai1998b,Kluzniak1998,Zhang2001,Dai2006,Metzger2011,Metzger2015,Lv2014}. The emission power of these GRBs is defined by the spin-down luminosity of the magnetar, which does not depend on the accretion rate, and hence, does not directly depend on the progenitor star properties. For binary LGRB progenitors, once the stars lose their partial envelopes caused by the binary interactions, the final pre-SN core structure is dramatically changed and even for massive progenitors ($\gtrsim 60~M_\odot$), a NS rather than a BH might be born \citep{Podsiadlowski2003}. In this case, BH hyperaccretion systems might not be suitable for LGRBs. The constraints discussed above would not apply for these systems.

\acknowledgments
This work was supported by the National Basic Research Program of China (973 Program) under grant 2014CB845800, the National Natural Science Foundation of China under grants 11473022 and 11573023.  AH was supported by an Australian Research Council (ARC) Future Fellowship (FT120100363) and NSF grant PHY-1430152 (JINA-CEE).

\clearpage

\end{document}